\documentclass[fleqn,reqno,a4paper,12pt]{amsart}

\usepackage[utf8]{inputenc}
\usepackage{amsmath}
\usepackage{amsfonts}
\usepackage{amssymb}
\usepackage[left=2cm,top=2cm,right=2cm,bottom=2cm]{geometry}

\begin{document}

\title[A bi-Hamiltonian Integrable Two Component Generalization of ...]{A bi-Hamiltonian Integrable Two Component Generalization of the third-Order Burgers Equation}

\dedicatory{Dedicated to Professor Yavuz Nutku who passed away on December 8, 2010.}

\author[D. Talati]{Daryoush Talati}
\address{Department of Engineering Physics, Ankara University 06100 Tando\u{g}an, Ankara}
\email{talati@eng.ankara.edu.tr}
\author[R. Turhan]{Ref\.{i}k Turhan}
\email{turhan@eng.ankara.edu.tr}

\newtheorem{thm}{Theorem}[section]
\newtheorem{prop}{Proposition}

\begin{abstract}
We announce a new bi-Hamiltonian integrable two-component system admitting the scalar 3rd-order Burgers equation as a reduction.
\end{abstract}

\maketitle

\section{Introduction}

Bi-Hamiltonian evolutionary equations are those which can be embedded in a Magri scheme 
\begin{eqnarray*}
\begin{array}{l}
u_{t_{i}}=F_{i}[u]=\mathrm{K}G_{i}[u]=\mathrm{J}G_{i+1}[u], \;\; G_{i}=\mathrm{E}(h_{i}[u]),\;\;\;i=-1,0,1,2,3,\cdots
\end{array}
\end{eqnarray*}
constructed by two compatible Hamiltonian (skew-adjoint, Jacobi identity satisfying) operators $\mathrm{J}$ and $\mathrm{K}$. Hamiltonian operators are defined to be  compatible if their arbitrary linear combinations are Hamiltonian operators also. Here the differential functions $F_{i}[u]$ are the (characteristics of) symmetries and the dual objects $G_{i}$ are the conserved gradients which are Euler derivatives $\mathrm{E}(h_{i})$ of the conserved densities $h_{i}$.

Very recently \cite{DSK}, it is rigorously proved that for a given two compatible Hamiltonian operators, construction of partial Magri scheme $(F_{i}, G_{i}, h_{i})$, $i=-1,0,1,2,\cdots,N$ for some finite $N$ implies existence of remaining infinitely many of them, and therefore integrability of the equations $F_{i}$ in the sense of existence of infinitely many  conservation laws. 

For a given pair of compatible Hamiltonian operators, construction of the Magri scheme is quite a straight forward task to do. The inverse problem of finding an appropriate compatible pair of Hamiltonian operators in whose Magri scheme a given system exists, is a nontrivial one.

The Ablowitz-Kaup-Newell-Segur (AKNS)\cite{AKNS} system
\begin{equation*}
\left(
\begin{array}{l}
u_{t}\\
v_{t}
\end{array}\right)=
\left(
\begin{array}{cc} 
-u_{xx} +  2u^{2}v\\
v_{xx} - 2v^{2}u
\end{array}
\right),
\end{equation*}
can be considered as a 2-component generalization of the scalar Heat equation (HE) in (1+1) dimensions, since $u=0$ reduction of the system leaves the  scalar HE $v_{t}=v_{xx}$ behind. 

The second order HE cannot be written as a bi-Hamiltonian equation but the odd-order members $u_{t}=u_{(2n+1)x}$, $n=0,1,2,3,\cdots$, where $u_{nx}=\frac{\partial^{n} u}{\partial x^{n}}$, in the (autonomous) symmetry hierarchy 
$u_{t}=u_{nx}$, of the HE are bi-Hamiltonian equations as noted recently in \cite{SKW} and given in a related variables in which the first member ($x$-translation symmetry of the HE) appears  in the Riemann equation form in \cite{ON} (see Proposition~3.3 in \cite{TT1}). 

Motivated by the observation that the second-order scalar HE, despite not being  a bi-Hamiltonian equation, has a bi-Hamiltonian 2-component generalization, i.e. the AKNS system, and that the odd-order HE's are bi-Hamiltonian, 
we considered a class of third-order Burgers type two-component systems for integrability. We shall give the results of our classification elsewhere \cite{TT2}. But here, starting from an example system which is related to a known integrable system, we introduce a new completely integrable 3rd-oder Burgers type system whose bi-Hamiltonian structure we constructed too. 

\section{An example related to a known system}
Let us consider the following system
\begin{equation}
\left(
\begin{array}{l}
u_{t}\\
v_{t}
\end{array}\right)=
\left(
\begin{array}{cc} 
u_{xxx} + 3u_{xx}u + 3u_{x}^{2} + 3u^{2}u_{x} + v_{xx} + 2u_{x}v \\
v_{xxx} -3v_{xx}u + 6u_{xx}v + 3u_{x}v_{x} + 3u^{2}v_{x} + 2vv_{x}
\end{array}\label{ayrel}
\right),
\end{equation}
which reduces to the scalar 3rd-order Burgers equation in $v=0$. In terms of the new variables $(w,z)$ which are defined 
as 
\begin{equation*}
w=u_{x},\;\;\; z=v+\frac{3}{2}u^{2} 
\end{equation*}
the above system becomes
\begin{equation}
\left(
\begin{array}{l}
w_{t}\\
z_{t}
\end{array}\right)=
\left(
\begin{array}{cc} 
w_{xxx} +z_{xxx} + 2w_{x}z+2wz_{x} \\
z_{xxx} -9ww_{x} + 6zw_{x} + 3wz_{x} + 2zz_{x}
\end{array}
\right). \label{aysys}
\end{equation}
The transformed system (up to a linear change of variables) is the Karasu (Kalkanl\i) equation which is obtained in Painlav\'{e} classification \cite{AyK}. A recursion operator is given for the system in \cite{AAS} and its bi-Hamiltonian structure is constructed by Sergyeyev in \cite{ArS}.

\section{The new system}

The new Burgers type 2-component system we introduce here is
\begin{equation}
\left(
\begin{array}{l}
u_{t}\\
v_{t}
\end{array}\right)=
\left(
\begin{array}{cc} 
u_{xxx} + 3u_{xx}u + 3u_x^2 + 3u_{x}u^2 + v_{xx} + u_{x}v+uv_{x}  \\
v_{xxx} -3v_{xx}u+3u_{xx}v+6uu_{x}v +3u^{2}v_{x}+4vv_{x}
\end{array}
\right).\label{sys}
\end{equation}
This system reduces to the 3rd-order Burgers equation by $v=0$ also. It can be written as 
a bi-Hamiltonian system as given in the following proposition.

\begin{prop} The system (\ref{sys}) is bi-Hamiltonian

\begin{equation*}
\left(
\begin{array}{cc} 
u_{t}\\
v_{t}
\end{array}
\right)=\mathrm{K}
\left(
\begin{array}{cc} 
\delta_{u}\\
\delta_{v}
\end{array}
\right)
\int \frac{v}{2} \; \mathrm{d}x=\mathrm{J}
\left(
\begin{array}{cc} 
\delta_{u}\\
\delta_{v}
\end{array}
\right)
\int \left(vu_{x}+\frac{1}{2}u^{2}v+\frac{1}{3}v^{2}\right) \mathrm{d}x
\end{equation*}
with the compatible pair of Hamiltonian operators 
\begin{equation*}
\mathrm{J}= \left(
\begin{array}{cc} 
-\frac{1}{3}D & D^{2}+2Du\\
-D^{2}+2uD & 4vD + 2v_{x}
\end{array}\right),\;\;\;
\mathrm{K}= \left(
\begin{array}{cc} 
\mathrm{K}_1&\mathrm{K}_2\\
\mathrm{K}_3&\mathrm{K}_4
\end{array}\right),
\end{equation*}
where
\begin{equation*}
\begin{array}{ll}
\mathrm{K}_1=&-D^3 +u^2D+uu_{x} - u_{x} D^{-1} u_{x}  ,   \\
\mathrm{K}_2=&D^{4}  + 4u D^{3}  + (9u_x +3v + 5u^2)D^{2} + (7u_{xx}+5v_{x}+16uu_x+3uv+2u^3)D \\
&+ (2u_{xxx} + 6u_{xx}u + 6u_{x}^{2} + 6u^{2}u_{x} + 2v_{xx} + 3u_{x}v + 2uv_{x}) - u_{x} D^{-1} v_{x}, \\
\mathrm{K}_3=&-\mathrm{K}^{*}_{2},\\
\mathrm{K}_4=&6vD^{3}+9v_{x}D^{2}+(7v_{xx}+12u_{x}v-12v_{x}u+12u^{2}v+9v^{2})D\\
&+(2v_{xxx}-6v_{xx}u+6u_{xx}v+12uu_{x}v+6u^{2}v_{x}+9vv_{x})-v_{x}D^{-1}v_{x},
\end{array}
\end{equation*}
and $D^{n}=\frac{\mathrm{d}^{n}\phantom{x}}{\mathrm{d}x^{n}}$, $(*)$ denotes the formal adjoint.
\end{prop}
A {\bf proof} follows from direct (and tedious) calculation that for arbitrary functions $\xi(x)$, $\eta(x)$ and
linear combinations $\mathrm{H}_{\lambda}=\mathrm{K}+\lambda \mathrm{J}$ with constant $\lambda$, the functional trivector 
\begin{equation*}
\Psi=prV_{\mathrm{H}_{\lambda}{\tiny \left( \begin{array}{l}\xi \\ \eta  \end{array} \right)}}
\left(\int \left(\xi \;\;\;\eta\right)\wedge \mathrm{H}_{\lambda}
\left(
\begin{array}{c} 
\xi\\
\eta
\end{array}\right)\mathrm{d}x\right)
\end{equation*}
vanishes independently from the value of $\lambda$, where $prV_{Q}$ is the prolongation of the vector field with characteristic $Q[u]$ \cite{OLV}.

The first few conserved densities of the hierarchy which are obtained first by the software in \cite{HR}, are listed below. 
\begin{equation*}
\begin{array}{l}
h_{-1}=u\\
h_{0}=\frac{v}{2}\\
h_{1}=vu_{x}+\frac{1}{2}u^{2}v+\frac{1}{3}v^{2}\\
h_{2}=-2u_{xx}v_{x} + vu_{x}^{2} - 6uu_{x}v_{x} + 3v^{2}u_{x} + 8u^{2}vu_{x}-\frac{4}{3}v_{x}^{2} +u^{4}v+2u^{2}v^{2}+\frac{7}{9}v^{3}\\
\vdots
\end{array}
\end{equation*}
These densities suffice to write a partial Magri scheme proving integrability of the system (\ref{sys}).

\section{Discussion}

Even though both of the systems  (\ref{ayrel}) and (\ref{sys}) reduce to the same 3rd-order Burgers equation by $v=0$, 
the Hopf-Cole transform $u\rightarrow u_{x}/u$ of the system (\ref{ayrel}) is no more a local evolutionary equation. 
With the same transformation, system  (\ref{sys}) transforms to a non-polynomial system having $v=0$ reduction to the scalar 3rd-order HE.

Most of the classified third order systems are generalizations of the KdV equation or equations related to it \cite{AF,SI1,MaF,GK} and the references therein.  System (\ref{sys}) cannot be transformed to a form having a reduction to KdV or to a related equation without some additional constraint by differential substitutions.

In the scalar case, the Hopf-Cole transformation relating the Heat and the Burgers equations transforms the conserved gradients of the odd-oder HE's into some nonlocal objects. But in the 2-component case, Burgers forms of the systems seems more suitable to be written in Hamiltonian form.
 
System generalizations of the Burgers equation are classified mainly by generalizing the 2nd-order scalar equation \cite{MSS,SI2,OS,MF,SanWa,TW} the references therein.
They are expected and turn out to possess symmetries at all orders like their scalar version.
However in the system  we gave here, 
similar to the other known Magri schemes, the integrable equations are only in odd orders and there are no (local) symmetries with even order. The odd-order scalar Heat hierarchy seems an exception to this general rule: The even-order Heat equations persist to be the symmetries of the odd-order ones, even if the only place we can put them is the position of conserved gradients in the Magri scheme. 

If a multi-component system is to be named according to the scalar equation it reduces by some field reductions, the Karasu (Kalkanl\i) system is not of KdV type since in its original form (\ref{aysys}), it does not reduce to KdV. 
However the related system (\ref{ayrel}) we gave here is genuinely Burgers type because it has a reduction to the 3rd-order scalar Burgers equation without any extra constraint left.

\section*{Acknowledgements}
D.T. is supported by TUBITAK PhD Fellowship for Foreign Citizens.

\end{document}